\documentclass{article}
\usepackage[T1]{fontenc}
\usepackage{wrapfig}
\usepackage[portuges,english]{babel}
\usepackage{amssymb,latexsym,amsmath,color,mathrsfs,ifsym,graphics,stmaryrd} 
\usepackage[colorlinks,linkcolor=blue,urlcolor=blue,citecolor=black,
plainpages=false,pdfpagelabels,breaklinks]{hyperref}
\usepackage{dialogue}
\usepackage{graphicx}
\title{{\sc Quantum Theory Needs No `Interpretation'\\But `Theoretical Formal-Conceptual Unity'\\
\smallskip
\smallskip
(Or: Escaping Ad\'an Cabello's ``Map of Madness''\\With the Help of David Deutsch's Explanations)}}

\author{{\sc Christian de Ronde}\thanks{Fellow Independent Researcher of the {\it Consejo Nacional de Investigaciones Cient\'{\i}ficas y T\'ecnicas} (CONICET).}}
\date{}

\usepackage[margin=2.05 cm]{geometry}

\begin{document}
\maketitle

\begin{center}
\begin{small}
Philosophy Institute Dr. A. Korn, Buenos Aires University - CONICET\\ 
Engineering Institute - National University Arturo Jauretche, Argentina\\
Federal University of Santa Catarina, Brazil.\\ 
Center Leo Apostel fot Interdisciplinary Studies, Brussels Free University, Belgium \\
\end{small}
\end{center}

\bigskip

\begin{abstract}
\noindent In the year 2000, in a paper titled {\it Quantum Theory Needs No `Interpretation'}, Chris Fuchs and Asher Peres presented a series of instrumentalist arguments against the role played by `interpretations' in QM. Since then ---quite regardless of the publication of this paper--- the number of interpretations has experienced a continuous growth constituting what Ad\'an Cabello has characterized as a ``map of madness''. In this work, we discuss the reasons behind this dangerous fragmentation in understanding and provide new arguments against the need of interpretations in QM which ---opposite to those of Fuchs and Peres---  are derived from a representational realist understanding of theories ---grounded in the writings of Einstein, Heisenberg and Pauli. Furthermore, we will argue that there are reasons to believe that the creation of  `interpretations' for the theory of quanta has functioned as a trap designed by anti-realists in order to imprison realists in a labyrinth with no exit. Taking as a standpoint the critical analysis by David Deutsch to the anti-realist understanding of physics, we attempt to address the references and roles played by `theory' and `observation'. In this respect, we will argue that the key to escape the anti-realist trap of interpretation is to recognize that ---as Einstein told Heisenberg almost one century ago--- {\it it is only the theory which can tell you what can be observed.} Finally, we will conclude that what QM needs is not a new interpretation but instead, a theoretical (formal-conceptual) consistent, coherent and unified scheme which allows us to understand what the theory is really talking about.
\end{abstract}

\renewenvironment{enumerate}{\begin{list}{}{\rm \labelwidth 0mm
\leftmargin 0mm}} {\end{list}}

\newcommand{\ita}{\textit}
\newcommand{\mcal}{\mathcal}
\newcommand{\mfrak}{\mathfrak}
\newcommand{\mbb}{\mathbb}
\newcommand{\mrm}{\mathrm}
\newcommand{\msf}{\mathsf}
\newcommand{\mscr}{\mathscr}
\newcommand{\lra}{\leftrightarrow}
\renewenvironment{enumerate}{\begin{list}{}{\rm \labelwidth 0mm
\leftmargin 5mm}} {\end{list}}

\newtheorem{dfn}{\sc{Definition}}[section]
\newtheorem{thm}{\sc{Theorem}}[section]
\newtheorem{lem}{\sc{Lemma}}[section]
\newtheorem{cor}[thm]{\sc{Corollary}}
\newcommand{\Proof}{\textit{Proof:} \,}
\newcommand{\cqd}{{\rule{.70ex}{2ex}} \medskip}

\bigskip

\bigskip

\bigskip

\section*{Introduction}

Today, the interpretation of Quantum Mechanics (QM) is not a subject that calls the attention of physicists. There are very few of them willing to take a public stance towards the problem of interpretation by raising a flag and fighting another. While some physicists prefer to remain at safe distance from the heated discussions of philosophers, most of them do not even know about the existence of these battles between interpretations. As Maximilian Schlosshauer \cite[p. 59]{Schlosshauer11} has recently described: ``It is no secret that a shut-up-and-calculate mentality pervades classrooms everywhere. How many physics students will ever hear their professor mention that there's such a queer thing as different interpretations of the very theory they're learning about? I have no representative data to answer this question, but I suspect the percentage of such students would hardly exceed the single-digit range.'' However, there is a very good reason why physicists do not pay attention to the interpretational debate about QM which populates today the journals and meetings of philosophy of science, and it is quite simple: what is considered by physicists to be `quantum theory' already provides all they need, namely, the prediction of `clicks' in detectors or `spots' in photographic plates. Thus, contemporary physicists seem to regard `interpretations' not so much as descriptions of reality but rather, as fictional stories created by armchair philosophers who tend to dream about imaginary worlds beyond empirical science.  

In this work we attempt to address the question of interpretation from a critical perspective. We will argue that, contrary to the orthodox claims according to which `interpretations' are an essential element of the realist quest, its function has been to trap the realist in a labyrinth with no exit. The paper is organized as follows. We begin by providing a representational realist account of physical theories. We then discuss the deep anti-realist re-foundation of physics that took place during the 20th Century. We then continue to show how this new understanding of theories was applied to the mathematical formalism of QM ---mainly, through the addition of the {\it projection postulate}--- producing what is orthodoxly known today as `Textbook' or `Standard' QM. From this standpoint, we address the problem of interpretation in quantum theory by considering the arguments presented by Fuchs and Peres. Taking into consideration the work by the physicist and philosopher David Deutsch, we criticize the orthodox understanding of theories and the role of interpretations in both physics and philosophy. Finally, we discuss a possible way out of this interpretational conundrum through the return to the original Greek understanding of physics supplemented by the insights produced in the work of Einstein, Heisenberg and Pauli.

\section{Physical Theories as Formal-Conceptual Representations}

For more than two millennia, since its Greek origin during the 7th Century B.C. up to the 16th Century in Modern times, physics was always related to {\it physis} ---`reality' or `Nature' as it was later on also translated (see \cite{Cordero14}). That is the reason why it was called `physics'.  {\it Physis}, the fundament of the whole existence, is the kernel concept ---grounding the {\it praxis} of both physics and philosophy--- which allowed to justify the search for unified theoretical schemes of thought that would explain phenomena. The main idea professed by the first philosophers was that reality possessed an internal order, what the Greeks called ---after Heraclitus--- a {\it logos}, and that this {\it logos} could be known through the development of {\it theories}. In order to learn about this logos the first philosophers begun the creation of {\it theories}, representational schemes about reality which were determined by a deep inter-relation between mathematics and concepts. Indeed, the first theories conceived the fundamental elements of {\it physis} as intrinsically related to mathematical forms. Since everyone could ---in principle--- learn about a reality that was not ruled by the wishes and desires of the Gods, physics ---which was also concomitant to the creation of Greek democracy--- became the most powerful enemy of the authority based knowledge that prevailed at the time. The many self-chosen figures like kings, mediums, priests and preachers, who had justified their ruling in societies in a preferred access to the divine had to give up part of their power to scientific understanding. Like democracy, both physics and philosophy flourished in the beautiful islands of the Aegean sea turning its capital, Athens, into the center of the Ancient world. This was until, a few Centuries later, the first form of anti-realism arrived to Athens. 

During the 5th. Century B.C., sophists ---as they became to be known--- produced the first attack against realism, fighting the idea that theoretical knowledge about {\it physis} (or reality) was possible. Sophists reframed the meaning of {\it logos} ---which philosophers had applied to {\it physis}--- as a linguistic form of discourse ---detached from any direct relation to {\it physis}--- and realized that its knowledge had an important pragmatic use, namely, to win debates in the Agora quite independently of being right or wrong. But sophists did not only sell their pragmatic knowledge ---devised to manipulate discourse through rhetorical tricks--- in the streets of Athens. They also criticized in depth the attempt to refer to reality. According to them, there is no such thing as `a reality of things'; and even if it existed, we would not be able to grasp it. We can only refer to our own perception. As Protagoras would argue, in a phrase that would become famously used by positivists of the {\it Vienna Circle}, ``Man is the measure of all things, of the things that are, that they are, of the things that are not, that they are not'' [DK 80B1]. We, individuals, have only a {\it relative} understanding of things dependent on our own experience. Taking a sceptic position, sophists criticized the physical idea of theoretical knowledge: realists are too naive, they believe they can access reality, but no one can. The first philosophers had already shown a recognition of this problem. Even though, there is an affinity between the {\it logos} of men and the {\it logos} of {\it physis}, it is a difficult task to expose the true {\it logos} since, as remarked by Heraclitus, ``{\it physis} loves to hide.'' [f. 123 DK]. Doing so requires hard work and sensibility, but ---following Heraclitus--- the latter can be revealed in the former. In a particular {\it logos} one can ``listen'' something that exceeds it, that is not only that personal discourse but the logos of {\it physis}: ``Listening not to me but to the {\it logos} it is wise to agree that all things are one'' [f. 50 DK]. We are thus able to represent {\it physis}, to exhibit its {\it logos}, but this representation should be not confused as mirroring reality. 

Since the first encounter between physicists and sophists, the battles between realists and anti-realists have never ceased in the history of Western thought. The attack of sophists was powerful and determined, showing the many lacunas of pre-Socratic realism. It took the strength of both Plato and Aristotle to control the damage. In their strive to answer sophists, Plato and Aristotle had to develop the first systematic categorical schemes of concepts, something that would become to be known with the term `metaphysics'. These conceptual architectures devised from first interrelated principles was able to answer, at least partly, the questions and problems that anti-realists had created in order to debunk the realist scheme. It is in this way that for two millennia scientific knowledge, through both Plato and Aristotle, prevailed in the west. Which does not mean that inverting the democratic role of science, many groups ---specially religious--- tried, and to great extent succeeded, in concentrating this knowledge for themselves ---hiding it from the pepole. For many centuries realism ruled developing knowledge and understanding of nature. This was until, during the 16th and 17th Century, anti-realism would begin to create a new order. One in which the subject would play an essential role. English empiricism jointly followed by rationalism begun the construction of a new foundation ---grounded on the subject--- for a new form of knowledge that would finally detach physics from {\it physis}.  


\section{The Anti-Realist Abduction of Physics: Theories as `Predictive Tools'}

After a long history of many lost battles anti-realists were finally able to prevail against their archenemy. As pointed out by David Deutsch \cite[p. 313]{Deutsch04}: ``During the twentieth century, anti-realism became almost universal among philosophers, and common among scientists. Some denied that the physical world exists at all, and most felt obliged to admit that, even if it does, science has no access to it.'' The unity of the anti-realist program was not provided by a common goal but ---instead--- by the common fear against realists. After Immanuel Kant, everyone was forced to agree that human {\it representations} were essentially incapable to grasp ---what the philosopher from K\"onigsberg had called--- {\it reality in itself} (in german, {\it Das ding an sich}). Since representations were necessarily dependent on the human categorical shaping, the possibility to link the foundation of knowledge in reality had became to be regarded as a metaphysical chimera of the past. We humans are finite, we cannot reach the infinite. In the context of science, it was mainly logical positivism which raised once again the flags of both sophistry and empiricism in order to re-found the meaning and reference of `physical theories' beyond both metaphysics and {\it physis}. With a different mask, these same ideas were also supported and reinforced by Niels Bohr, one of the most influential figures in the anti-realist development of physics during the first half of the 20th Century \cite{deRonde20}. After the Second World War, all these lines finally converged in a much more explicit and radical form of anti-realism, namely, U.S. instrumentalism. However, regardless of their minor differences, what is essential to notice is that there exists a common structure to all forms of anti-realism which is rooted in sophistry ---as acknowledged by the Vienna Circle in their famous manifesto--- and grounded on two main pillars. First, the idea that subjective perception is the fundamental ground of any meaningful statement. Second, that any discourse is {\it relative} to a community of subjects, not to an ``external reality'' ---as anti-realist would re-name {\it physis}. It is at this point that language, the way in which a community of subjects were able to communicate and reach consensus, became to be regarded as the new foundation of knowledge.\footnote{This foundation is of course also related to what has been called in both continental and analytic traditions the ``linguistic turn''.}
\begin{enumerate}
{\bf  \item[I.] Naive Empiricism:} Observation is a self evident {\it given} of ``common sense'' experience. 
{\bf  \item[II.] Intersubjective Relativism:} The {\it intersubjective} communication and consensus reached by the members of a scientific community allows to regard a theory as being ``objective''. Objective does not mean anymore the provision of a consistent categorical {\it moment of unity} but ---instead--- an ``honest'' way to collect data and share it with other members of the community.
\end{enumerate}
The effective strategy of anti-realists was grounded on the creation of an ambiguous dictionary in which realist words were detached from their meaning and reference. As stressed by Deutsch: ``In some fields (such as statistical analysis) the very word `explanation' has come to mean prediction, so that a mathematical formula is said to `explain' a set of experimental data. By `reality' is meant merely the observed data that the formula is supposed to approximate. That leaves no term for assertions about reality itself, except perhaps `useful fiction'.'' `Objectivity' which implied the possibility of constituting a conceptual {\it moment of unity} within a subject-independent representation was also replaced ---by Bohr--- towards the `inter-subjective' communication of empirical data. And while `truth' was now understood as approximate and partial, `metaphysics' would become to be regarded as a mere narrative or interpretation.
\begin{enumerate}
{\bf \item[III.] Anti-Foundational:} Theories do not provide a true reference to {\it physis} (or reality). Theories are always provisional and ungrounded.   
{\bf \item[IV.] Anti-Metaphysical:} Metaphysics, understood as an interpretation (or narrative) about {\it unobservable entities}, is not essentially required within empirically adequate theories ---it plays no essential role within physics. 
\end{enumerate}
During the 20th Century, physicists were taught to become agnostic about matters concerning reality. The notion of `reality' which was responsible for the very creation of scientific thought and the first fight against religious and mystic storytelling was now reframed as a religious belief. According to anti-realists, instead of making reference to reality, theories begin with the observation of phenomena and end with their prediction. Period. Science was re-named as ``empirical science''  and physical theories became to be regarded as algorithmic predictive devices about observable measurement outcomes. The radical skepticism learned by the new generations of physicists and philosophers helped them understand there was nothing to be found beyond observable phenomena ---only illusions. All discourse added beyond observation was to be regarded as metaphysical bla bla. Or, as physicists and philosophers say today, ``just a way of talking''. A theory only provides a partial, temporary success which ---in the long run--- will certainly show its failures and will be then replaced by another new theory to come. Physics is not about {\it physis}, it is just a language designed to communicate  observational data between the members of a scientific community. As explained by Niels Bohr \cite{Bohr60}: ``Physics is to be regarded not so much as the study of something a priori given, but rather as the development of methods of ordering and surveying human experience. In this respect our task must be to account for such experience in a manner independent of individual subjective judgement and therefor objective in the sense that it can be unambiguously communicated in ordinary human language.'' The only reality that exists is that which can be observed. But there is no foundation grounding such observations. As Karl Popper would famously remark:  
\begin{quotation}
\noindent {\small ``The empirical basis of objective science has thus nothing `absolute' about it. Science does not rest upon solid bedrock. The bold structure of its theories rises, as it were, above a swamp. It is like a building erected on piles. The piles are driven down from above into the swamp, but not down to any natural or `given' base; and if we stop driving the piles deeper, it is not because we have reached firm ground. We simply stop when we are satisfied that the piles are firm enough to carry the structure, at least for the time being.'' \cite[p. 111]{Popper92}} 
\end{quotation}

\section{The Anti-Realist Foundation of Standard Quantum Mechanics}

Quite regardless of the fact that most physics professors might accept ``off the record'' that ``nobody understands QM'' ---a phrase made popular by the U.S. physicist Richard Feynman---, there is nonetheless a widespread unspoken consensus of how to teach the theory to students. This can be witnessed from the fact that all graduate textbooks present the theory more or less on the same lines, as a mathematical formalism capable through a series of rules to compute and predict measurement outcomes. This is the way in which QM is taught to students in all Universities around the world, namely, as a ``recipe''. As explained by Tim Maudlin:
\begin{quotation}
\noindent {\small ``What is presented in the average physics textbook, what students learn and researchers use, turns out not to be a precise physical theory at all. It is rather a very effective and accurate recipe for making certain sorts of predictions. What physics students learn is how to use the recipe. For all practical purposes, when designing microchips and predicting the outcomes of experiments, this ability suffices. But if a physics student happens to be unsatisfied with just learning these mathematical techniques for making predictions and asks instead what the theory claims about the physical world, she or he is likely to be met with a canonical response: Shut up and calculate!'' \cite[pp. 2-3]{Maudlin19}} 
\end{quotation}
The origin of this orthodox presentation of QM, also known as ``Standard QM'', can be traced back to the publication during the first years of the 1930s of two books which attempted to produce an axiomatic formulation of the theory of quanta in terms of operators on a Hilbert space \cite{Dirac74, VN}. Both of them were  clearly aligned not only with Bohr's interpretation of QM but also with the positivist (anti-realist) understanding of physical theories. 

Since its origin, the mathematical formulation of QM was surrounded by controversy. It was in year 1925 that Werner Heisenberg was finally able to present a sound mathematical formulation of the theory of quanta ---a task he finished with the help of the mathematicians Pascual Jordan and Max Born. The non-continuous mathematical representation together with its highly abstract rules horrified the physicists of the time, accustomed to handle differential equations which could be framed in a spatiotemporal representation. As Einstein \cite[p. 195]{QTen} would write to his friend Michele Besso: ``The most interesting recent theoretical achievement, is the Heisenberg-Born-Jordan theory of quantum states. A real sorcerer's multiplication table, in which infinite determinants (matrices) replace Cartesian coordinates. It is extremely ingenious, and thanks to its great complication sufficiently protected against disproof.'' But only one year later, Erwin Schr\"odinger's wave formulation gave a new hope to a conservative community of physicists still expecting to restore a ``more classical'' representation of atoms and their trajectories through space and time ---something that Heisenberg had explicitly left behind in order to reach matrix mechanics. However, regardless of his efforts, it was soon acknowledged by Schr\"odinger himself that the hopes were ungrounded. The quantum wave function, $\Psi$, could not be understood as a spatiotemporal entity. Its mathematical domain was not 3-dimensional but {\it configuration space}, a multi-dimensional mathematical space in which the number of degrees of freedom was relative to the number of wave-functions considered. Paul Dirac, a young English engineer and mathematician, was observing the problem from a different angle. Dirac simply wanted to provide a sound mathematical presentation of the theory that would reproduce ---what he considered to be--- observations in the lab. After all, following positivism, physical theories were to be understood just as mathematical schemes capable to account for observations. That was it. As he stressed in the introduction of his book, {\it The Principles of Quantum Mechanics}, it is ``important to remember that science is concerned only with observable things [...]''. Dirac also realized immediately not only the importance of quantum superpositions but the essential difficulty which the superposition principle implied for such an empirical-positivist understanding of theories. The problem was not that quantum superpositions were difficult to represent conceptually for as Dirac has  remarked, ``the main object of physical science is not the provision of pictures, but the formulation of laws governing phenomena and the application of these laws to the discovery of phenomena. If a picture exists, so much the better; but whether a picture exists or not is a matter of only secondary importance.'' The problem was that such quantum superpositions introduced a serious obstacle for his positivist reading in terms of the certain prediction of definite observations. The theory simply did not make reference to the sudden appearance of single `clicks' in detectors; instead, it described weird superposed states evolving and even interacting in a realm in which multiple possibilities seemed to be taking place simultaneously ---something that both Einstein and Schr\"odinger would make explicit five years later. In order to bridge this gap, making use of an example of polarized photons, Dirac introduced for the first time the now famous ``collapse'' of the quantum wave function:
\begin{quotation}
\noindent {\small ``When we make the photon meet a tourmaline crystal, we are subjecting it to an observation. We are observing wither it is polarized parallel or perpendicular to the optic axis. The effect of making this observation is to force the photon entirely into the state of parallel or entirely into the state of perpendicular polarization. It has to make a sudden jump from being partly in each of these two states to being entirely in one or the other of them. Which of the two states it will jump cannot be predicted, but is governed only by probability laws.'' \cite[p. 7]{Dirac74}} 
\end{quotation} 
Two years later, the already famous Hungarian mathematician John von Neumann, took the empirical-positivist program a step further. By turning the collapse of the quantum superposition into an {\it axiom}, von Neumann was able to present Dirac's {\it ad hoc} jumping rule as part of the theory itself. In his own \cite[p. 214]{VN} words: ``Therefore, if the system is initially found in a state in which the values of $\mathcal{R}$ cannot be predicted with certainty, then this state is transformed by a measurement $M$ of $\mathcal{R}$ into another state: namely, into one in which the value of $\mathcal{R}$ is uniquely determined. Moreover, the new state, in which $M$ places the system, depends not only on the arrangement of $M$, but also on the result of $M$ (which could not be predicted causally in the original state) ---because the value of $\mathcal{R}$ in the new state must actually be equal to this $M$-result'' In this way, Dirac's quantum jump was converted from being an arbitrary introduced rule into a decent measurement postulate of the theory. The book by von Neumann, {\it Mathematical Foundations of Quantum Mechanics}, ``became really the Bible of the so called Copenhagen interpretation of QM'' \cite{Andrei08}, another name which physicists have been using ever since in order to describe the Standard ``recipe'' of QM.

\section{The Measurement Problem and the Call for Interpretations}

As remarked by Klas Landsman \cite[p. 435]{Landsman17}: ``The measurement problem of quantum mechanics was probably born in 1926'', the reason being that in a famous paper published that year, Max Born proposed his now famous probabilistic interpretation of Schr\"odinger's quantum wave function introducing a direct reference to measurement outcomes making an implicit use ---already then--- of the ``collapse'' of the quantum wave function. However, it was not Born but Dirac and von Neumann who would make completely explicit the reference to this new ``quantum jump'' ---between the mathematical representation of quantum superpositions and the observation of single outcomes--- introduced within the theory. The inconsistencies of this proposal were very soon acknowledged by Albert Einstein who discussed them with Sch\"odinger in many different occasions. Both of them, in complete solitude, repeatedly insisted in the serious problems which the introduction of measurement implied within the theory. In 1935 their combined attack against the anti-realist account of QM took place through a series of papers which would attempt ---in vain--- to re-frame the theory of quanta in terms of a consistent definition of physical reality. Einstein wrote to Schr\"odinger after his publication of today's famous `cat paper': 
\begin{quotation}
\noindent {\small You are the only contemporary physicist, besides Laue, who sees that one cannot get around the assumption of reality, if only one is honest. Most of them simply do not see what sort of risky game they are playing with reality ---reality as something independent of what is experimentally established. They somehow believe that the quantum theory provides a description of reality, and even a complete description; this interpretation is, however, refuted most elegantly by your system of radioactive atom + Geiger counter + amplifier + charge of gun powder + cat in a box, in which the $\Psi$-function of the system contains both the cat alive and blown to bits. Nobody really doubts that the presence or absence of the cat is something independent of the act of observation.'' \cite[p. 39]{ESPL}} 
\end{quotation} 
Their realist offensive was immediately controlled by Niels Bohr himself, the commander of the anti-realist troops, who used all his rhetorical powers to dissolve their questioning. Bohr's arguments were elusive and not very convincing. As a matter of fact, it took several decades for the same community of physicists who vigorously claimed that Bohr had finally settled the matter to realize that there were two pages of his published paper in the incorrect order. As mentioned by David Albert \cite{Albert19}: ``[T]he standard version of Bohr's response to the EPR, the one that was almost exclusively the template for every reprinting it, reprinting of it over a period of about 50 years, had two of the pages reversed.'' In 1935, Bohr was crowned as the new champion and congratulated by a physicist community already tired of Einstein's ``too philosophical'' considerations. And then the war started...

After the Second World War, the triumph of anti-realism was finally complete and the criticisms of two rebels were eventually forgotten. The Bohrian-positivist unspoken alliance converged then into a new pragmatic trend of thought even more radical than its predecessors. Instrumentalism was presented by the the U.S. philosopher John Dewey as the natural extension of both pragmatism and empirical positivism. Pragmatism sustained that the value of an idea is determined by its usefulness, but instrumentalism ---rejecting the need of any metaphysical fundament--- was ready to take a step further and claim that the question regarding the {\it reference} of theories was simply meaningless. {\bf Scientific theories do not make reference to an underlying reality.} Thus, there is no sense in which a theory could be said to be {\it true} or {\it false} (or better or worse) apart from the extent to which it is useful as a ``tool'' in solving scientific problems. As described by Olival Freire Jr.:  
\begin{quotation}
\noindent {\small ``In the US, which after the Second World War became the central stage of research in physics in the West, the discussions about the interpretation of quantum mechanics had never been very popular. A common academic policy was to gather theoreticians and experimentalists to gather in order to favour experiments and concrete applications, rather than abstract speculations. This practical attitude was further increased by the impressive development of physics between the 1930s and the 1950s, driven on the one hand by the need to apply the new quantum theory to a wide range of atomic and subatomic phenomena, and on the other hand by the pursuit of military goals. As pointed out by Kaiser, `the pedagogical requirements entailed by the sudden exponential growth in graduate student numbers during the cold war reinforced a particular instrumentalist approach to physics'.'' \cite[pp. 77-78]{Freire15}} 
\end{quotation} 
By the 1960s, the new orthodoxy was firmly in power and the reference to {\it physis} had been erased completely from the understanding of theories. As famously described by Karl Popper \cite{Popper63}:  ``Today the view of physical science founded by Osiander, Cardinal Bellarmino, and Bishop Berkeley, has won the battle without another shot being fired. Without any further debate over the philosophical issue, without producing any new argument, the {\it instrumentalist} view (as I shall call it) has become an accepted dogma. It may well now be called the `official view' of physical theory since it is accepted by most of our leading theorists of physics (although neither by Einstein nor by Schr\"odinger). And it has become part of the current teaching of physics.'' Thus, while physical theories became to be regarded as `tools' used by agents in order to compute measurement outcomes, questions about physical reality became enclosed as purely ``philosophical debates''. The technical application of QM became the main ambition of instrumentalist physicists, and the bombs they created, their justification for the need of a completely new pragmatic understanding of science. As Popper would remark: ``Instead of results due to the principle of complementarity other and more practical results of atomic theory were obtained, some of them with a big bang. No doubt physicists were perfectly right in interpreting these successful applications as corroborating their theories. But strangely enough they took them as confirming the instrumentalist creed.''

Instrumentalism had conquered physics, but realists were very slowly beginning to regroup. During the mid-1960s the EPR paper was finally rediscovered. Two theorems, one by John Bell and another by Simon Kochen and Ernst Specker, were developed in order to address ---once again--- the mathematical representation of QM in relation to physical reality. In this context, philosophy of physics would become a shelter for many banished physicists who were not ready to give up on the notion of reality. In this new context realists would begin to discuss in the corridors, when no one was listening, new and old narratives about the underlying message of the theory of quanta. This is what ---in more technical terms---  became to be known as an ``interpretation'' of the theory. As explained by Meinard Kuhlmann and Wolfgang Pietsch: 
\begin{quotation}
\noindent {\small ``[A] main task for philosophers of physics is the formulation and evaluation of suitable ontologies for specific physical theories. Since ---in contrast to general philosophy--- these ontologies are always tailored towards a given physical theory, they are often called interpretations. If `interpretation' is understood in this specific way, it can be characterized more exactly as a mapping of certain elements of the given theory to entities (e.g. particles, properties, structures) in the world, to which the theory is supposed to refer.''  \cite[p. 210]{KuhlmannPietsch12}} 
\end{quotation} 

More and more realists were willing to openly discuss about the reality of quanta. But taking note of the situation, anti-realist were also there to recall realists an essential lesson from the now orthodox anti-realists understanding of physics. Theories make no reference to anything beyond observable phenomena. In order to control the realist uprising, the interpretations of QM were immediately redirected towards the measurement problem which ---as we have seen--- originated itself from anti-realist presuppositions and requirements. As Craig Callender \cite[p. 385]{Callender09} explains, ``[t]he metaphysics of quantum mechanics [...] hangs on both a particular solution to the measurement problem and then the best interpretation of that solution.'' In this way, the infamous (anti-realist) measurement problem remained at the center of the stage of all (realist) philosophical debates about QM. As Tim Maudlin explains today: ``The most pressing problem today is the same as ever it was: to clearly articulate the exact physical content of all proposed `interpretations' of the quantum formalism is commonly called the measurement problem, although, as Philip Pearle has rightly noted, it is rather a `reality problem'.'' One might wonder why anti-realists had allowed some philosophers to address realist questions and metaphysical problems. But the truth is that even within instrumentalism the reference to a reality beyond observation was never completely abandoned. The reason is simple: without a representational language it becomes simply impossible to describe any experience or measurement. As stressed by Deutsch \cite{Deutsch04}: ``Scientific explanations are about reality, most of which does not consist of anyone's experiences.''  Paleontology talks about dinosaurs, not about the observation of fossils; biology talks about living creatures, not about moving spots in microscopes; and physics has always talked about interacting entities like particles (in classical mechanics) and waves (in electromagnetism), not about `clicks' in detectors or `spots' in photographic plates. Without a representational account of a states of affairs it becomes impossible to do science. But 20th Century anti-realists had learned the lesson of Mach's defeat in the hands of the atom. Thus, they had allowed some unwanted metaphysical references that went far beyond empirical observations. Atoms ---which Mach had fought against just a few decades before--- were accepted as part of a new discourse about QM. Elementary quantum particles, could be regarded as being responsible for our macroscopic immediate experience about chairs and tables, even though in the end they were ``just a way of talking''. Of course, the addition of interpretations to the theory was considered by anti-realists as something completely different. 

Anti-realists allowed the debate about interpretations within philosophy of physics, but they did so not without diminishing its role as a fictional drug consumed by realist philosophers in order to calm their metaphysical anxieties. Metaphysical representational discourse was an unwanted guest that had to be tolerated. Postmoderns had learned that when controlled, a few metaphysical bed time stories made no harm to anyone. The miracle is that, contrary to all expectations, the weakness created by this inconsistent double role of representational language ---on the one hand as a metaphysical device considered as a fiction, and on the other, as making reference to our ``common sense'' or ``manifest image'' of the world--- was turned by anti-realists into the major strength of their system. Inconsistency was used as a rhetorical weapon in order to stop the realist attempt to produce a unified, consistent and coherent representation of QM. The following dialogue, which has repeated itself countless times in Universities all around the world, between a young student eager to understand Nature and a Professor in charge of an introductory course on QM shows how the trick works. With a dramatic emphasis the Professor begins his speech before teaching how to compute {\it eigenvectors} and {\it eigenvalues}.    
\begin{dialogue}
\speak{Professor} {\it Today we are beginning our course about QM. Our best physical theory ever!! 

QM talks about microscopic particles which are the constituting building-blocks of our macroscopic world. Most technological developments that have taken place in the 20th Century are related to the knowledge we have acquired about quantum particles.}

\smallskip

\speak{Student} {\it That is fascinating! 

So what are these quantum particles like? What are they?}

\smallskip

\speak{Professor} {\it What do you mean? What are they...?

Well, these particles are very, very small, so we cannot actually see them... we influence them every time we measure them by using aparatuses... but, in fact, these quantum particles are also waves ---depending, of course, on the measurement we choose to perform. And what's really fascinating is that they don't even exist if we do not observe them.} 

\smallskip

\speak{Student} {\it Wait a moment... What do you mean? The particles are waves? They only exist when observed?} 

\smallskip

\speak{Professor} {\it I already told you that they are ``quantum'' particles! 

Of course QM does not talk about `particles'. How can you think something like that? 

This is all just a way of talking...} 

\smallskip

\speak{Student} {\it Sorry. But I'm not sure I understand... A ``way of talking''?} 

\smallskip

\speak{Professor} {\it Well, the point is that nobody really understands what QM is taking about... But don't worry. QM does predict the correct measurement outcomes with an incredible accuracy, so that's it.  

And what are these questions suppose to mean anyhow? 

Why don't you better... Shut up and calculate!}
\end{dialogue}

One might regard this dialogue as simply a bad or even inconsistent way of teaching physics. But maybe the major strength of postmodern anti-realism presents itself in the form of inconsistency, the fact it is able to support both anti-realist and realist claims simultaneously, depending on the context o inquiry. This pendular rhetorical game is very common between physicists who ---like our Professor--- always begin by arguing that QM provides a (realist) reference to a microscopic realm constituted by elementary particles, but when asked about this realm, immediately shift the reference to the (anti-realist) prediction of measurement outcomes and pragmatic accomplishments. The measurement problem might be then regarded as a reflection of this same inconsistency: an attempt to provide a realist justification ---namely, an `interpretation'--- of an anti-realist rule ---introduced to the mathematical formalism in a completely {\it ad hoc} manner--- which was meant to save not only single measurement outcomes (or empirical phenomena) but also the anti-realist (mis)understanding of QM itself.

\section{Interpretations of Quantum Theory: A Map of Madness?}

Since the Second World War, physicists around the world have been trained ---for many decades now--- in an instrumentalist fashion, learning that the questions about reality (or {\it physis}) are of a purely philosophical nature, not at all part of physics.\footnote{As described by Arthur Fine: ``[The] instrumentalist moves, away from a realist construal of the emerging quantum theory, were given particular force by Bohr's so-called `philosophy of complementarity'; and this nonrealist position was consolidated at the time of the famous Solvay conference, in October of 1927, and is firmly in place today. Such quantum nonrealism is part of what every graduate physicist learns and practices. It is the conceptual backdrop to all the brilliant success in atomic, nuclear, and particle physics over the past fifty years. Physicists have learned to think about their theory in a highly nonrealist way, and doing just that has brought about the most marvelous predictive success in the history of science.'' \cite[p. 1195]{PS}} However, already by the end of the 1950s several physicists ---from previous generations--- who were still concerned with the missing reference of QM to physical reality begun to develop the first set of realist interpretations for the theory of quanta. An important group of them was going back to the almost forgotten Aristotelian (hylemorphic) metaphysical scheme: Heisenberg's and Simondon's potentiality interpretation ---later on developed in logical terms by Josef-Maria Jauch and Constantin Piron---, Margenau's latency and even Popper's propensity narratives; all of them had the same goal, namely, try to make sense of quantum superpositions and Born's rule (see \cite{deRonde17, Koznjak20}). David Bohm, going back to de Broglie's ideas,  would attempt to restore a classical actualist metaphysical representation in which quantum particles would possess definite trajectories. Of course, none of these proposals had any influence within a physics community which had already agreed on what Standard QM meant and was only interested in the practical success of its possible applications. In 1964, using his weekends,\footnote{As Bell \cite{Bell83} would ironically remark in the opening of his `underground colloquium': ``I am a Quantum Engineer, but on Sundays I Have Principles''.} John Bell ---an Irish physicist working at the CERN--- was going back to EPR's questioning by re-considering the meaning of physical reality in terms of classical probability. A few years later, in 1967, Simon Kochen and Ernst Specker, taking as a standpoint the quantum mathematical formalism itself, would discuss the consistent binary valuation of projection operators in different bases. All this was of course taking place during weekends, in the margins of the instrumentalist ruling.  However, a point of inflection ---in what was now considered the philosophical research of QM--- occurred when during the first years of the 1980s Alain Aspect's experiments, taking Bell's analysis as a standpoint, showed the strength of quantum correlations in the form of {\it quantum entanglement} ---the same forgotten notion which Einstein and Schr\"odinger had discussed in 1935.\footnote{Diederik Aerts recalls that by the late 1970s physicists did not know about the notion of quantum entanglement. Private correspondence, July 2020.} It was these series of experiments which gave a powerful impulse to the still young field of philosophy of QM. Slowly, through their analysis of EPR, hidden variables, contextuality and non-locality ---between many other subjects---, the new field of research was beginning to gain recognition. Suddenly, philosophical problems and questions which had been almost completely silenced since Bohr's triumph against Einstein were once again being discussed in the open. In this new context, an increasing number of researchers ---not only philosophers and physicists, but also logicians and mathematicians--- still interested in foundational questions started to gather in philosophical meetings about QM. A great example of this new phenomena is the series of encounters that took place in Johensu, Finland, in the years 1985, 1987 and 1990, organized by the physicist Kalervo Laurikainen. The three Symposia Proceedings of these meetings ---jointly edited with Claus Montonen, Pekka Lathi and Pieter Mittelstaedt--- show the passion of the discussions that were already then taking place. At this point, in order to find answers to forgotten questions, interpretations begun once again to be developed in different directions: neo-Bohmian, modal, informational, many worlds, many minds, consistent histories, instantaneous collapse, contextual, logical... Old and new interpretations could be now freely discussed. But of course, realists were not alone. Bas van Fraassen \cite[p. 10]{VF80}, one of the most prominent and influential anti-realist philosophers of physics, would also come to these meetings just to recall realists that ``theories need not be true to be good.'' And that `To develop an empiricist account of science is to depict it as involving a search for truth only about the empirical world, about what is actual and observable.'' As he would argue, an interpretation \cite[p. 242]{VF91} should respond to the question ``what would it be like for this theory to be true, and how could the world possibly be the way this theory says it is? [...] However we may answer these questions, believing in the theory being true or false is something of a different level.'' Van Fraassen \cite[p. 8]{VF80}, an anti-realist, would explain to realists their role as fanatic or naive believers: ``Science aims to give us, in its theories, a literally true story of what the world is like; and acceptance of a scientific theory involves the belief that it is true.'' In contraposition, he would picture himself and his fellow anti-realists companions as part of a more down to earth, agnostic movment: ``Anti-realism is a position according to which the aim of science can well be served without giving such a literally true story, and acceptance of a theory may properly involve something less (or other) than belief that it is true.'' The bottom line of van Fraassen's speech was meant to remind the realist that her interpretations were just that, fictions, narratives, stories with no link to the (empirical) theory. As a way to control the ongoing debate anti-realists ---regardless of the fact they were supposedly not interested in finding a representation of the theory beyond phenomena--- begun to introduce their own ``interpretations'', subverting in this way its precarious realist reference. Since then, anti-realist interpretations of QM have become as popular as the realist ones: van Fraassen's own modal interpretation, Everett's relative states interpretation, Bitbol's contextual interpretation, Dieks' perspectival interpretation, Fuch's QBism and Rovelli's informational interpretation are just a few of the many anti-realist interpretations that populate today contemporary journals. 

But Aspect's experiments did not only give an impulse to philosophy of QM. Maybe even more importantly, they also begun to crack the instrumentalist dogma from within physics itself. As described by Jeffrey Bub:
\begin{quotation}
\noindent {\small  ``[...] it was not until the 1980s that physicists, computer scientists, and cryptographers began to regard the non-local correlations of entangled quantum states as a new kind of non-classical resource that could be exploited, rather than an embarrassment to be explained away. [...] Most physicists attributed the puzzling features of entangled quantum states to Einstein's inappropriate `detached observer' view of physical theory, and regarded Bohr's reply to the EPR argument (Bohr, 1935) as vindicating the Copenhagen interpretation. This was unfortunate, because the study of entanglement was ignored for thirty years until John Bell's reconsideration of the EPR argument (Bell, 1964).'' \cite{Bub17}}
\end{quotation}
One decade later, during the 1990s, instrumentalist trained physicists  were beginning to develop a new field of research that would be known as {\it Foundations of Quantum Mechanics}. Then, something extraordinary happened. Even philosophers of QM were allowed to present their findings in the meetings of physicists. Physicists seemed to need help from philosophers! Quite ironically, in the heart of anti-realist physics, the new technical research of quantum information processing was stuck with the same (metaphysical) problems and questions that the two realist rebels ---Einstein and Schr\"odinger--- had produced more than half a century before. 

\begin{figure}
\centering
\includegraphics[scale=.4]{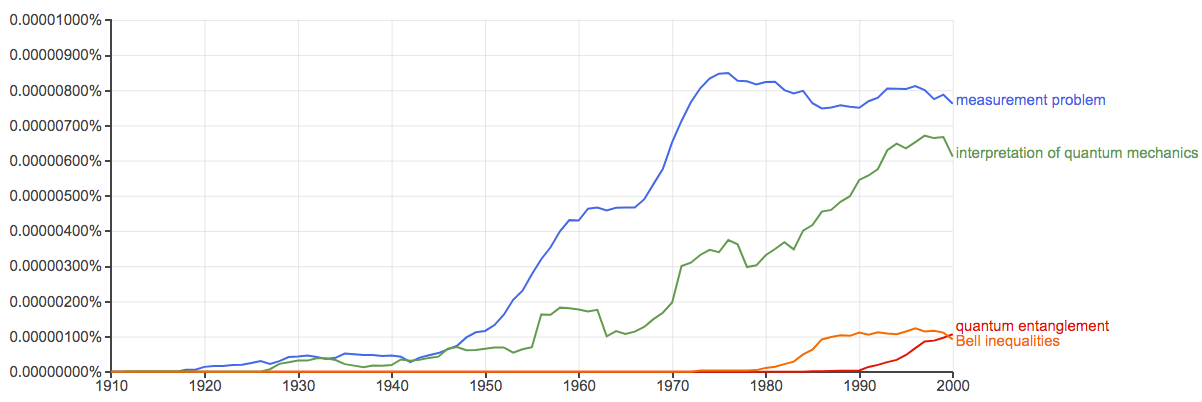}\caption{\small{The image from Google Ngram Viewer shows the relevance of the terms `measurement problem', `interpretation of quantum mechanics', `Bell inequalities' and `entanglement' present in books between the years 1910 and 2000.}}
\end{figure}

\smallskip

In the year 2000, exactly one century after the birth of the quantum, Chris Fuchs and Asher Peres wrote a paper titled {\it Quantum Theory Needs No Interpretation}, in which they provided a series of instrumentalist arguments against the need of producing an interpretation for QM. As they explicitly argued \cite[p. 70]{FuchsPeres00}: ``[...] quantum theory does not describe physical reality. What it does is provide an algorithm for computing probabilities for the macroscopic events (`detector clicks') that are the consequences of experimental interventions. This strict definition of the scope of quantum theory is the only interpretation ever needed, whether by experimenters or theorists.'' Both Fuchs and Peres were right to point out that, as already acknowledged in both physics and philosophy of physics, an empirically adequate theory ---such as QM--- was all that was actually needed for the theoretical or experimental physicist. Interpretations cannot change what has been already observed. As recognized even by Dennis Dieks, a promotor of the ---supposedly--- realist modal interpretation: ``an interpretation cannot change the empirical content and predictive power of the theory that is involved''. In short, Fuchs and Peres are absolutely correct to point out that the addition of interpretations is neither justified nor required. As Peres had already explained in the introduction of his famous book, {\it Quantum Theory: Concepts and Methods}: 
\begin{quotation}
\noindent {\small ``[S]ome physicists tend to attribute to the wave function $\Psi$  the objective status that was lost by $q$ and $p$. There is a temptation to believe that each particle (or system of particles) has a wave function, which is its objective property. [...] Unfortunately, there is no experimental evidence whatsoever to support this naive belief. On the contrary, if this view is taken seriously, it leads to many bizarre consequences, called `quantum paradoxes'. These so-called paradoxes originate solely from an incorrect interpretation of quantum theory. The latter is thoroughly pragmatic and, when correctly used, never yields two contradictory answers to a well posed question. It is only the misuse of quantum concepts, guided by a pseudorealistic philosophy, which leads to these paradoxical results.'' \cite[p. 4]{Peres93}}
\end{quotation}
The recommendation made by Arthur Fine \cite[p. 149]{Fine86} seems then quite reasonable: ``Try to take science on its own terms, and try not to read things into science. If one adopts this attitude, then the global interpretations, the `isms' of scientific philosophies, appear as idle overlays to science: not necessary, not warranted and, in the end, probably not even intelligible.'' 

 
Fuchs and Peres grounded their arguments against interpretations in today's orthodox understanding of physical theories, conceived as mathematical formalisms adequately supplemented by rules which allow to predict empirical observations. This essential definition is accepted not only by anti-realist philosophers, but ---more importantly--- also by realist ones. Notwithstanding this fact, the most commonly widespread justification of the need of `interpretations' is grounded on the realist claim that {\bf physical theories talk about reality}. And ---of course--- it is interpretations themselves which are supposed to carry the realist content of theories. This is the reason why interpretations ---which supposedly explain what reality is like according to the theory--- become essential for realists. Interpretations are the carriers of all realists hopes to make contact with reality. The problem is that this realist idea ---commonly stated not only by philosophers but also by physicists--- cannot be consistently considered in the context of the anti-realist present understanding of physics which states ---as explicitly acknowledged by Fuchs and Peres--- that exactly the opposite is true, namely, that {\bf physical theories do not make reference to reality.} According to the orthodox contemporary understanding, physical theories are only required ---as van Fraassen puts it--- to ``save the observed phenomena''. Period. Realists are faced then with an unsolvable problem, to show how their interpretations have some kind of objective link to the theory they ---supposedly--- refer to. Without such a link, anti-realists are correct to point out that these stories seem nothing but a fictional game outside the empirical scientific enterprise. 

The realist philosopher is compelled to present an objective justification for the addition of her interpretations. But is this possible to do so in the context of the anti-realist understanding of theories? There are serious obstacles which show that this realist project is doomed to failure. The first problem is that according to the anti-realist account of theories, different families of mathematical models can explain the same table of (observed) data  ---a possibility most famously exemplified in the Standard Model of elementary particles. Thus, it is always possible to create new theories from exactly the same set of observations.\footnote{The ongoing debate regarding the understanding of theories is between different anti-realists viewpoints such as the {\it syntactic}, the {\it semantic} and the {\it pragmatic} approaches all of which take for granted the idea that observations are unproblematic (see \cite{Winther16}).} The existence of this possibility, to account for the same table of data from different theories, gives rise to the so called {\it theoretical underdetermination} problem. If different mathematical schemes provide exactly the same predictions it becomes obviously impossible to choose between them on experimental grounds. In the context of the observation of quanta, Standard QM, Bohmian mechanics and GRW are good examples of this same problematic situation. All of these different mathematical formalisms (or theories) attempt to describe exactly the same phenomena, so how could we choose between them? The problem gets more complicated since each and every mathematical formalism can be also subject to different interpretations or stories about the underlying reality of the theory. Philosophers can create many different narratives to account for the same theory. But how can anyone choose between them? Which takes us to the problem of {\it interpretational underdetermination}. QM does not determine its interpretation. In fact, Standard QM can be interpreted in terms of many worlds, many minds, consistent histories, conscious collapses, potentialities, propensities, latencies, and the list continues... Bohmian mechanics and GRW also have many different interpretations in terms of propensities, flashes, quantum fields, etc.\footnote{At this point, we should remark that in the specialized literature there is even no consensus regarding the limits between `theory' and `interpretation'. The confusion between `theory' and `interpretation' in the specialized literature is explicit in QM where Bohmian mechanics and GRW are some times regarded as different `theories', and sometimes conceived as different `interpretations' of the same quantum theory. Also, interpretations are used by both realists and anti-realists to discuss theories. As remarked by Alejandro Cassini in \cite{Cassini17}, the term `interpretation' is not a well defined notion within philosophy of physics.} So how can a realist choose to believe in her preferred interpretation? This process of fragmentation in our understanding of experience does not stop here. Even worse, each and every interpretation can be subject to an ontological scheme which should make clear what are the {\it entities} to which the interpretation refers to. The same interpretation can be subject to different ontologies, different accounts of what these entities really are. There are many different ontological readings for each interpretation. Thus, we have reached what is called in the specialized literature {\it ontological underdetermination}. Finally, even if we could be able to justify our choice of a particular mathematical formalism, of a specific interpretation, and a particular ontology, we would still have to face what Steven French has called {\it metaphysical underdetermination} \cite{French11}. That is, the problem of making explicit the metaphysical scheme of the chosen ontology.\footnote{Regarding the distinction between metaphysics and ontology we might refer to Thomas Hofweber \cite[p. 13]{Hofweber16}: ``In metaphysics we want to find out what reality is like in a general way. One part of this will be to find out what the things or the stuff are that are part of reality. Another part of metaphysics will be to find out what these things, or this stuff, are like in general ways. Ontology, on this quite standard approach to metaphysics, is the first part of this project, i.e. it is the part of metaphysics that tries to find out what things make up reality. [. . . ] Ontology is generally carried out by asking questions about what there is or what exists.''} And once again, we are stuck with the same problem. There are no objective reasons why to choose one metaphysical proposal instead of another. At best, we might evaluate all these choices ---of formalisms, interpretations, ontologies and metaphysics--- only subjectively, in terms of aesthetic standards \cite{Benovsky16}, voluntarism \cite{Chakravartty17} or ---simply put--- in terms of ``personal preferences'' \cite{Maudlin19b} (see for a detailed analysis of these problems \cite{Raoni20}). 
\begin{enumerate}
{\bf  \item Theoretical (or Mathematical) Underdetermination:} There are many different mathematical models capable to account for the same table of observable data. So how can we choose between them?
{\bf \item Interpretational Underdetermination:} There are many different interpretations which can act as narratives about the underlying reality to a which a theory refers to. So how can we choose between these different made up stories?
{\bf \item Ontological Underdetermination:} There are many different ontologies which explain, in different ways what are the entities of an interpretation of a theory. So how can we choose between them? 
{\bf \item Metaphysical Underdetermination:} There are many different metaphysics which can account for the ontology of an interpretation of a theory. So how do we know what metaphysics is the true one?
\end{enumerate}


There is no single physicist nor philosopher who entering this labyrinth has been able to find a way out. The reason might have been all the time just in front of our eyes: This anti-realist scheme was never designed for realist purposes. Funny enough, it is the complete lack of anti-realist answers regarding the meaning of the theory of quanta which is probably also responsible for the increase of the ``realist narratives'' during the last decades. As David Mermin \cite[p. 8]{Mermin12} remarks: ``[Q]uantum theory is the most useful and powerful theory physicists have ever devised. Yet today, nearly 90 years after its formulation, disagreement about the meaning of the theory is stronger than ever. New interpretations appear every day. None ever disappear.'' So almost one century after the first mathematical formulation of QM in 1925, the overpopulated number of interpretations in the literature has configured what Ad\'an Cabello has characterized as ``A map of madness'' \cite{Cabello17}. From a philosophical perspective, one might regard this situation as a very precarious one, where fragmentation seems to hunt understanding. On the contrary, we tend to believe that this is the result of a very efficient anti-realist scheme designed to trap the realist in a labyrinth with no exit. The realist has been confined to perform a Promethean task: an endless meaningless search for narratives which have no objective justification, use, nor contact with the theory they attempt to refer to. Anti-realists seem to have applied in a very efficient way the strategy proposed by Michael Corleone in {\it The Godfather}: ``Keep your friends close, but your enemies closer.'' By providing realists something meaningless to work on, anti-realists have been finally able to control them... Maybe for the first time in the history of a war ---between physicists and sophists--- which begun more than two and a half millennia ago, anti-realists have been able to capture realists within their rhetorical net. Anti-realists have created a maze in which realists are free to wonder as fantastic preachers of made up stories that no one really cares about. But like in every conjuring trick the illusion can only work if the realist believes everything she has been told ---by the anti-realist--- is actually true. In this respect, it might be sad to acknowledge that most contemporary realists might have become comfortable with the role they've been given as naive or fanatic believers. Well, not everyone...


\section{David Deutsch's Lonely Fight Against Anti-Realism in Physics}

David Deutsch is maybe the only contemporary physicist who instead of accepting the anti-realist challenge to try to find an interpretation,\footnote{As remarked by Deutsch in a talk given to Everett : ``The term `interpretation' has come to have connotations that misrepresent not only quantum theory but science in general.''} has confronted the main premisses of anti-realism itself. There are three essential points made by Deutsch in his criticism to the anti-realist understanding of physics. First, the idea that theories are not derived from observations. Second, that theories cannot be grounded on intersubjective authority, not even coming from scientific communities. And third, that theories are not designed as predictive machineries but rather to provide `good explanations'. Let us discuss these points in some detail. 

Attacking the main cornerstone from which the whole anti-realist architectonic has been constructed, Deutsch argues against the idea that science is derived from observations.  
\begin{quotation}
\noindent {\small ``The fact that the light was emitted very far away and long ago, and that much more was happening there than just the emission of light ---those are not things that we see. We know them only from theory. Scientific theories are explanations: assertions about what is out there and how it behaves. Where do these theories come from? For most of the history of science, it was mistakenly believed that we `derive' them from the evidence of our senses ---a philosophical doctrine known as empiricism.'' \cite[pp. 3-4]{Deutsch04}}
\end{quotation}
We certainly agree with the main point raised by Deutsch. However, it is also important ---at least, for our purposes--- to correct an essential aspect of it: the idea that knowledge is derived from the senses is not original form English empiricism, it goes back to Greek sophistry. It is sophists, not empiricists, who first argued that the only accessible knowledge that humans could gain came from their own sense-perception. The anti-realist conclusion was that (human) knowledge is {\it relative} to (human) perception and thus, the idea of (external) reality is nothing but a fiction, a chimera invented by physicists and philosophers. Sophists were the firsts to question ---during the V Century B.C.--- with very good and precise arguments, the foundation of physics itself ---namely, {\it physis} (or reality). As we argued above, it took the strength of both Plato and Aristotle to control the sophist attack. In order to overcome the anti-realist uprising they were forced to develop the first systematic metaphysical schemes of thought; i.e., categorical architectonics of interrelated concepts. In fact, the reference to sophists was explicitly embraced by logical (or empirical) positivists who wrote in their famous manifesto: ``Everything is accessible to man; and man is the measure of all things. Here is an affinity with the Sophists, not with the Platonists; with the Epicureans, not with the Pythagoreans; with all those who stand for earthly being and the here and now.'' But let us continue. According to Deutsch, observations are always {\it theory-laden}.\footnote{As Deutsch \cite[p. 9-10]{Deutsch04} remarks: ``empiricists came to believe that, in addition to rejecting ancient authority and tradition, scientists should suppress or ignore any new ideas they might have, except those that had been properly `derived' from experience. As Arthur Conan Doyle's fictional detective Sherlock Holmes put it in the short story `A Scandal in Bohemia', `It is a capital mistake to theorize before one has data.' But that was itself a capital mistake. We never know any data before interpreting it through theories. All observations are, as Popper put it, theory-laden,* and hence fallible, as all our theories are.''}
\begin{quotation}
\noindent {\small ``scientific theories are not `derived' from anything. We do not read them in nature, nor does nature write them into us. They are guesses ---bold conjectures. Human minds create them by rearranging, combining, altering and adding to existing ideas with the intention of improving upon them. We do not begin with `white paper' at birth, but with inborn expectations and intentions and an innate ability to improve upon them using thought and experience. Experience is indeed essential to science, but its role is different from that supposed by empiricism. It is not the source from which theories are derived. Its main use is to choose between theories that have already been guessed. That is what `learning from experience' is.'' \cite[p. 4]{Deutsch04}}
\end{quotation}

With respect to authority, today's experts have become the new mediums, the priests of a postmodern era in which only `specialists' can have a voice, while the rest of mortals can only expect to ``understand'' science through made up stories ---which in fact, as recognized by specialists themselves, have nothing to do with the theory they attempt to explain. It has become accepted that only specialists can talk about their field of expertise, and only they can understand what they themselves say. This applies not only to the layman but also to other experts from different sub-fields. In physics, for example, it is accepted that a researcher in one sub-field should not necessarily understand a single word of a presentation given by a colleague working in a different sub-field. A physicists working in particle physics is not supposed to understand a physicist working in cosmological theories ---and, of course, vice versa. As a trained physicist, I myself have experienced in too many occasions the unpleasant feeling of not being able to understand absolutely anything after the first slide of a talk. As a consequence, just like in the interpretations of QM, scientific debates become fragmented in different sub-fields of research, each one of them restricted to a limited group of experts. David Deutsch has characterized all this situation simply as ``bad philosophy''. 
\begin{quotation}
\noindent {\small ``Bad philosophy has always existed too. For instance, children have always been told, `Because I say so.' Although that is not always intended as a philosophical position, it is worth analysing it as one, for in four simple words it contains remarkably many themes of false and bad philosophy. First, it is a perfect example of bad explanation: it could be used to `explain' anything. Second, one way it achieves that status is by addressing only the form of the question and not the substance: it is about who said something, not what they said. That is the opposite of truth-seeking. Third, it reinterprets a request for true explanation (why should something-or-other be as it is?) as a request for justification (what entitles you to assert that it is so?), which is the justified-true-belief chimera. Fourth, it confuses the nonexistent authority for ideas with human authority (power) ---a much-travelled path in bad political philosophy. And, fifth, it claims by this means to stand outside the jurisdiction of normal criticism.''  \cite[p. 311]{Deutsch04}}
\end{quotation}

Finally, we reach the third point made by Deutsch.  It is {\it explanation}, not interpretations, which ---he argues--- should be regarded as the kernel constituent of scientific theories. But instead of linking this aspect of theories to the ancient Greek tradition, Deutsch prefers to relate it to 17th Century Enlightenment. 
\begin{quotation}
\noindent {\small ``The quest for good explanations is, I believe, the basic regulating principle not only of science, but of the Enlightenment generally. It is the feature that distinguishes those approaches to knowledge from all others, and it implies all those other conditions for scientific progress I have discussed: It trivially implies that prediction alone is insufficient. Somewhat less trivially, it leads to the rejection of authority, because if we adopt a theory on authority, that means that we would also have accepted a range of different theories on authority. And hence it also implies the need for a tradition of criticism. It also implies a methodological rule ---a criterion for reality--- namely that we should conclude that a particular thing is real if and only if it figures in our best explanation of something.'' \cite[p. 22-23]{Deutsch04}}
\end{quotation}
By making explicit something difficult to deny ---namely, that theories make reference to situations or states of affairs---, Deustch \cite[p. 6]{Deutsch04} continues to undermine the anti-realist reference to subjective experience:  ``Scientific explanations are about reality, most of which does not consist of anyone's experiences. Astrophysics is not primarily about us (what we shall see if we look at the sky), but about what stars are: their composition and what makes them shine, and how they formed, and the universal laws of physics under which that happened. Most of that has never been observed: no one has experienced a billion years, or a light year; no one could have been present at the Big Bang; no one will ever touch a law of physics ---except in their minds, through theory. All our predictions of how things will look are deduced from such explanations of how things are.'' The same can be said of Paleontology \cite{Deutsch11}: ``We never speak of the existence of dinosaurs millions of years ago as an interpretation of our best theory of fossils. We say, or, we claim that it is the explanation of fossils. And what's more, that it is not primarily about fossils it is about dinosaurs.'' Continuing his critical evaluation Deutsch creates a deeply interesting analogy between the anti-realist understanding of theories and conjuring tricks. Using this analogy, he makes explicitly clear the way in which anti-realists have been able to subvert the meaning and content of some of the most essential notions of the realist understanding of theories:  
\begin{quotation}
\noindent {\small ``Some people may enjoy conjuring tricks without ever wanting to know how they work. Similarly, during the twentieth century, most philosophers, and many scientists, took the view that science is incapable of discovering anything about reality. Starting from empiricism, they drew the inevitable conclusion (which would nevertheless have horrified
the early empiricists) that science cannot validly do more than predict
the outcomes of observations, and that it should never purport to
describe the reality that brings those outcomes about. This is known as
instrumentalism. It denies that what I have been calling `explanation'
can exist at all. It is still very influential. In some fields (such as statistical
analysis) the very word `explanation' has come to mean prediction, so
that a mathematical formula is said to `explain' a set of experimental
data. By `reality' is meant merely the observed data that the formula is
supposed to approximate. That leaves no term for assertions about
reality itself, except perhaps `useful fiction'. Instrumentalism is one of many ways of denying realism, the commonsense,
and true, doctrine that the physical world really exists, and is
accessible to rational inquiry. Once one has denied this, the logical
implication is that all claims about reality are equivalent to myths,
none of them being better than the others in any objective sense. That
is relativism, the doctrine that statements in a given field cannot be
objectively true or false: at most they can be judged so relative to some
cultural or other arbitrary standard. Instrumentalism, even aside from the philosophical enormity of reducing science to a collection of statements about human experiences,
does not make sense in its own terms. For there is no such thing as a
purely predictive, explanationless theory. One cannot make even the
simplest prediction without invoking quite a sophisticated explanatory
framework.'' \cite[p. 15]{Deutsch04}}
\end{quotation}

To sum up, let us conclude this section with a quite long but essential passage from Deutsch's book, {\it The Beginning of Infinity}, where he explains with great clarity how we reached this untenable situation. 
\begin{quotation}
\noindent {\small ``During the twentieth century, anti-realism became almost universal among philosophers, and common among scientists. Some denied that the physical world exists at all, and most felt obliged to admit that, even if it does, science has no access to it. For example, in `Reflections on my Critics' the philosopher Thomas Kuhn wrote: {\it There is [a step] which many philosophers of science wish to take and which I refuse. They wish, that is, to compare [scientific] theories as representations of nature, as statements about `what is really out there'.} Positivism degenerated into logical positivism, which held that statements not verifiable by observation are not only worthless but meaningless. This doctrine threatened to sweep away not only explanatory scientific knowledge but the whole of philosophy. In particular: logical positivism itself is a philosophical theory, and it cannot be verified by observation; hence it asserts its own meaninglessness (as well as that of all other philosophy). The logical positivists tried to rescue their theory from that implication (for instance by calling it `logical', as distinct from philosophical), but in vain. Then Wittgenstein embraced the implication and declared all philosophy, including his own, to be meaningless. He advocated remaining silent about philosophical problems, and, although he never attempted to live up to that aspiration, he was hailed by many as one of the greatest geniuses of the twentieth century.

One might have thought that this would be the nadir of philosophical thinking but unfortunately there were greater depths to plumb. During the second half of the twentieth century, mainstream philosophy lost contact with, and interest in, trying to understand science as it was actually being done, or how it should be done. Following Wittgenstein, the predominant school of philosophy for a while was `linguistic philosophy', whose defining tenet was that what seem to be philosophical problems are actually just puzzles about how words are used in everyday life, and that philosophers can meaningfully study only that. Next, in a related trend that originated in the European Enlightenment but spread all over the West, many philosophers moved away from trying to understand anything. They actively attacked the idea not only of explanation and reality, but of truth, and of reason. Merely to criticize such attacks for being self-contradictory like logical positivism ---which they were--- is to give them far too much credence. For at least the logical positivists and Wittgenstein were interested in making a distinction between what does and does not make sense ---albeit that they advocated a hopelessly wrong one.

One currently influential philosophical movement goes under various names such as postmodernism, deconstructionism and structuralism, depending on historical details that are unimportant here. It claims that because all ideas, including scientific theories, are conjectural and impossible to justify, they are essentially arbitrary: they are no more than stories, known in this context as `narratives'. Mixing extreme cultural relativism with other forms of anti-realism, it regards objective truth and falsity, as well as reality and knowledge of reality, as mere conventional forms of words that stand for an idea's being endorsed by a designated group of people such as an elite or consensus, or by a fashion or other arbitrary authority. And it regards science and the Enlightenment as no more than one such fashion, and the objective knowledge claimed by science as an arrogant cultural conceit. Perhaps inevitably, these charges are true of postmodernism itself: it is a narrative that resists rational criticism or improvement, precisely because it rejects all criticism as mere narrative. Creating a successful postmodernist theory is indeed purely a matter of meeting the criteria of the postmodernist community ---which have evolved to be complex, exclusive and authority-based. Nothing like that is true of rational ways of thinking: creating a good explanation is hard not because of what anyone has decided, but because there is an objective reality that does not meet anyone's prior expectations, including those of authorities. The creators of bad explanations such as myths are indeed just making things up. But the method of seeking good explanations creates an engagement with reality, not only in science, but in good philosophy too ---which is why it works, and why it is the antithesis of concocting stories to meet made-up criteria.

Although there have been signs of improvement since the late twentieth century, one legacy of empiricism that continues to cause confusion, and has opened the door to a great deal of bad philosophy, is the idea that it is possible to split a scientific theory into its predictive rules of thumb on the one hand and its assertions about reality (sometimes known as its `interpretation') on the other. This does not make sense, because ---as with conjuring tricks--- without an explanation it is impossible to recognize the circumstances under which a rule of thumb is supposed to apply. And it especially does not make sense in fundamental physics, because the predicted outcome of an observation is itself an unobserved physical process.'' \cite[pp. 313-315]{Deutsch04}}
\end{quotation}

We certainly agree with the points made by Deutsch. However, we also believe that the term `explanation' requires a much deeper and careful characterization. One which should make clear the deep inter-relations between: theory, representation, thought-experience, observation, measurement and reality (see \cite{deRonde20}). It is these relations which ---going back to the ancient Greeks--- we attempt to address in the final section of this work.



\section{Theoretical Representational Unity and Quantum Mechanics}

Science originated as a non-authoritarian discipline in which, maybe for the first time in the history of humanity, theoretical representation allowed to avoid super-natural pseudo-explanations in terms of the actions of almighty Gods and other divinities. Good explanations, trying to escape mere subjective opinion ({\it doxa}), became the main goal of the first philosophers: Anaxagoras, Tales, Democritus, Heraclitus and Parmenides ---between many others--- begun to provide different theories about {\it physis} (or reality) providing explanations of natural phenomena \cite{Cordero14}. As Jean-Paul Vernant makes explicitly clear, the revolution that begun with the Milesians in Greece was one of enormous proportions. One that determined the origin of science itself. 
\begin{quotation} 
\noindent {\small ``[Before the Milesians,] [e]ducation was based not on reading written texts but on listening to poetic songs transmitted from generation to generation. [...] These songs contained everything a Greek had to know about man and his past ---the exploits of heroes long past; about the gods, their families, and their genealogies, about the world, its form, and its origin. In this respect, the work of the Milesians is indeed a radical innovation. Neither singers nor poets nor storytellers, they express themselves in prose, in written texts whose aim is not to unravel a narrative thread in the long line of a tradition but to present an explanatory theory concerning certain natural phenomena and the organization of the cosmos. In this shift from the oral to the written, from the poetic song to prose, from narration to explanation, the change of register corresponds to an entirely new type of investigation ---new both in terms of its object (nature, {\it physis}) and in terms of the entirely positive form of thought manifested in it.'' \cite[p. 402]{Vernant06}} 
\end{quotation}

Scientific understanding was the first to confront the deep influence of powerful figures who claimed within different societies to have access to what would happen in the future. Priests, mediums, preachers and gurus justified their knowledge through the use of metaphors, studied ambiguity, and pseudo-explanations which, apart from themselves, no one else was expected to understand. On the very contrary, since its origin physics was always related to the democratic production of consistent, coherent and unified schemes of thought capable to comprise experience and phenomena in theories, producing in this way true understanding about {\it physis} (or reality). Everyone should be capable of understanding a scientific theory. As Pauli would explain to a very young Heisenberg:   
\begin{quotation}
\noindent {\small  ``knowledge cannot be gained by understanding an isolated phenomenon or a single group of phenomena, even if one discovers some order in them. It comes from the recognition that a wealth of experiential facts are interconnected and can therefore be reduced to a common principle. [...] `Understanding' probably means nothing more than having whatever ideas and concepts are needed to recognize that a great many different phenomena are part of coherent whole. Our mind becomes less puzzled once we have recognized that a special, apparently confused situation is merely a special case of something wider, that as a result it can be formulated much more simply. The reduction of a colorful variety of phenomena to a general and simple principle, or, as the Greeks would have put it, the reduction of the many to the one, is precisely what we mean by `understanding'. The ability to predict is often the consequence of understanding, of having the right concepts, but is not identical with `understanding'.'' \cite[p. 63]{Heis71}}
\end{quotation}
Physics is all about consistent and coherent theoretical unity. Conceptual unity, namely, the consistent interrelation of a specific set of concepts makes possible the creation of notions (e.g., `particles', `fields', `waves') which allows us to represent experience and phenomena beyond change and becoming. Mathematical unity allows us to compute experience through a consistent and coherent rigorous language. Theoretical unity, as the consistent conjunction of both formal and conceptual representations, allows us to create systems capable to account for thought-experience as well as {\it hic et nunc} observations in both a qualitative and a quantitative manner. 

Of course, as any program, the realist search for consistent and coherent theoretical unified schemes of thought, faces specific obstacles. There are two main problems which realists must confront with respect to coherency, consistency and unity. The first problem is the relation between {\it physis} and theories. What is the specific way in which theories relate to reality? Until the revolution that took place in the 16th and 17th centuries, both Plato and Aristotle were central in the many debates that took place in both physics and metaphysics. This was at least until Newtonian physics took its place as a theory capable to capture the true essence of the Universe in a spatio-temporal frame. Every question that man had investigated for centuries was now answered in a systematic manner in terms of a representation which provided both qualitative and quantitive understanding of phenomena. In the 17th Century, physics seemed to have finally achieved its goal of representing reality. However, at this point, philosophy produced a counter-revolution in which the justification of true physical knowledge was severely questioned. After Hume and Kant, the widespread naive understanding of Newtonian mechanics as mirroring reality became completely untenable. Since then, the relation between theoretical representation and reality became one of the main sources of controversy. The orthodox idea according to which theories should describe reality ({\it transcendent relation}) implies an unspoken reference to a final representation ---still waiting to be unveiled, the {\it Theory of Everything} (T.O.E.). This, of course, places the realist in a very week position.\footnote{This idea has been strongly supported by empirical scientists like Stephen Hawking, Steven Weinberg, Gerard t'Hooft and Michio Kaku, between many others.} Elsewhere, we have discussed a different possibility grounded on an {\it immanent relation}, much closer to the Greek immanent understanding of theories as inherent expressions of reality \cite{deRonde16b}. From this realist perspective, a theory need not mirror reality. Representation does not bring anything into presence, it is always more than that which presents itself. Representation is not a mirror, it might be better thought as a conceptual net, a machine. Theoretical representations are designed to capture through their contact with the here and now, singular expressions of reality. Theories operate {\it within} reality, they are created from within. What is real? The singular experience described by a theory which takes place through measurement in the lab {\it hic et nunc}. Or in other words, the coherent and consistent theoretical representation of experience which takes place during actual measurements in the lab as described by the theory. Every time we use a consistent closed theory in order to produce and understand a measurement outcome, we can say that reality has been expressed in an adequate manner (see for a more detailed analysis \cite[Sect. 5]{deRonde20}). It is neither the representation nor the measurement outcome which is real, what is real is the is the singular expression of reality captured in between theory and actual experience through measurement. Each and every different theoretical net captures what it has been designed to apprehend. While classical mechanics cannot capture an expression of an electromagnetic field, Maxwell's theory will not be able to describe the bouncing of two particles. What is real? The coming together of theoretical representation and {\it hic et nunc} experience in a given actual situation. This understanding of the relation between theories and reality implies that theoretical representation is not necessarily unique. There can exist many different representations, many different nets capable to capture different expressions of reality. While, representation is not committed to description, realism is not committed to the search of a single representation of reality. The search for our understanding of reality can be regarded as an infinite creative process of theories which express reality in an adequate manner. 

The second problem is the specific relation between theory and experience ---something that anti-realists have been completely unable to explain within their scheme. For realists, as Hume and Kant taught us, it is only the theory which decides what can be observed. Metaphysical concepts as conditions of possibility for theoretical thinking play an essential role which cannot be reduced to a mere narrative. As remarked by Einstein: 
\begin{quotation}
\noindent {\small``From Hume Kant had learned that there are concepts (as, for example, that of causal connection), which play a dominating role in our thinking, and which, nevertheless, can not be deduced by means of a logical process from the empirically given (a fact which several empiricists recognize, it is true, but seem always again to forget). What justifies the use of such concepts? Suppose he had replied in this sense: Thinking is necessary in order to understand the empirically given, {\it and concepts and `categories' are necessary as indispensable elements of thinking.}'' \cite[p. 678]{Einstein65} (emphasis in the original)}
\end{quotation} 
Only the theory, and in particular the concepts it uses, can explain what has been observed. As Heisenberg made the point \cite[p. 264]{Heis73}: ``The history of physics is not only a sequence of experimental discoveries and observations, followed by their mathematical description; it is also a history of concepts. For an understanding of the phenomena the first condition is the introduction of adequate concepts. Only with the help of correct concepts can we really know what has been observed.'' What is observed depends on the concepts used by each different theory. Since every theory talks in terms of different concepts and formalisms, there is in principle no limit, no bridge to be found between them. The concepts applied by each theory imply not ``just a way of talking'' but {\it a way of thinking}. Each and every physical theory must be a closed formal-conceptual unified scheme of thought capable to express a specific field of phenomena. While concepts are required in order to represent experience in a qualitative fashion, mathematical formalisms are required in order to compute experience in a quantitative manner. The interrelation between conceptual schemes and mathematical formalisms takes place through two essential notions: invariance and objectivity. These two notions act as mutual counterparts in the construction of theoretical representations which, as a precondition for coherency and consistency, must be independent of reference frames and subjective perspectives. While {\it operational-invariance} allows us to grasp in a mathematical formalism what can be considered to be {\it the same}, independently of a reference frame; {\it objectivity} allows us to do something analogous in conceptual terms, creating conceptual {\it moments of unity} (e.g., particles, waves, fields, etc.) which are independent of subjective observations. In order to catch reality there can be no holes in the net. Both conceptual and mathematical schemes must weave their notions in a  coherent and consistent manner, something which also applies for their inter-relation. A theory must act as a formal-conceptual-operational consistent, coherent and unified whole. As remarked by Heisenberg \cite[p. 95]{Bokulich06}: ``One finds [in closed theories] structures so linked and entangled with each other that it is really impossible to make further changes at any point without calling all the connections into question [...] We are reminded here of the artistic ribbon decorations of an Arab mosque, in which so many symmetries are realized all at once that it would be impossible to alter a single leaf without crucially disturbing the connection of the whole.'' As a consequence of this unity between concepts and mathematical formalisms, one should be able to go from the concepts to the mathematics ---like it was the case in classical mechanics--- and vice-versa, from the mathematics to the concepts ---as it seems to be required in the case of QM. This essential requirement of theoretical unity, not only avoids ---right from the start--- the possibility to introduce a measurement postulate ---grounded on a naive empirical prejudice about single measurement outcomes---, it also erases many interpretations which add concepts to the theory which have no contact to the mathematical formalism whatsoever ---most of them grounded on metaphysical prejudices.\footnote{As Chris Fuchs \cite[p. 71]{Schlosshauer11} remarks: ``Take the nearly empty imagery of the many-worlds interpretation(s). Who could derive the specific structure of complex Hilbert space out of it if one didn't already know the formalism?'' The same can be stated of Bohmian mechanics where there is no mathematical representation neither of particles nor their classical trajectories.}  

To conclude, given these general lines of what realism actually means, the following question rises: What is the problem of QM from a realist perspective of analysis? The answer can be found when going back to the original meaning of physics. Physics provides its own guiding lines, its own methodology grounded on the attempt to produce a consistent and coherent theoretical formal-conceptual unified scheme in which all pieces of the puzzle fit together in such a manner that no piece can be extracted without destroying the whole system. This is something very difficult to achieve and has been developed in very specific occasions through the hard work of many different communities. The consistency must not only interrelate mathematics and concepts, it must be also able to connect thought and actual experience. According to this analysis, the problem of QM is to find a unified conceptual scheme which makes contact with the mathematical formalism in a manner in which we are able to produce an invariant-objective representation of a state of affairs. What we need is an operational-invariant mathematical formalism capable to account in a quantitative manner for experimental tests and an objective (subject-independent) conceptual representation which is capable to make a qualitative sense of experience in a coherent and consistent manner. Is this realist representation possible in the context of QM? We believe it is, for as we have learned from the history of scientific developments ``a new explanation is inherently an act of creativity'' \cite[p. 7]{Deutsch04}.

\section*{Acknowledgements} 

I want to thank Raimundo Fern\'andez Mouj\'an, Mat\'ias Graffigna, Juan Vila and Raoni Wohnrath Arroyo for many discussions on related subjects. This work was partially supported by the following grants: FWO project G.0405.08 and FWO-research community W0.030.06. CONICET RES. 4541-12 (2013-2014).


\begin{thebibliography}{9999}

\bibitem{Albert19} Albert, D., 2019, ``David Albert on Quantum Measurement and the Problems with Many-Worlds'', {\it Sean Carroll's Mindscape Podcast (Episode 36)}, https://www.preposterousuniverse.com

\bibitem{Bell83} Bell, J., 1983, `underground colloquium' from March 1983.

\bibitem{Benovsky16} Benovsky, J.,  2016, {\it Meta-metaphysics: On metaphysical equivalence, primitiveness, and theory choice.} Springer, Switzerland. 

\bibitem{Bohr60} Bohr, N., 1960, {\it The Unity of Human Knowledge},
In {\it Philosophical writings of Neils Bohr}, vol. 3., Ox Bow
Press, Woodbridge.

\bibitem{Bokulich06} Bokulich, A., 2006, ``Heisenberg Meets Kuhn:
Closed Theories and Paradigms'', {\it Philosophy of Science}, {\bf
73}, 90-107.

\bibitem{Bub17} Bub, J., 2017, ``Quantum Entanglement and Information'', {\it The Stanford Encyclopedia of Philosophy (Spring 2017 Edition)}, Edward N. Zalta (ed.), URL = https://plato.stanford.edu/archives/spr2017/entries/qt-entangle/.

\bibitem{Cabello17} Cabello, A., 2017, ``Interpretations of quantum theory: A map of madness'', in {\it What is Quantum Information?}, pp. 138-143,  O. Lombardi, S. Fortin, F. Holik and C. L\'opez (eds.), Cambridge University Press, Cambridge.

\bibitem{Callender09} Callender, 2009, ``Metaphysics of Quantum Mechanics'', in {\it Compendium of Quantum Physics: Concepts, Experiments, History and Philosophy} D. Greenberger and K. Hentschel and F. Weinert (Eds.), Springer, Berlin.

\bibitem{VC} Carnap, H., Hahn, H. $\&$ Neurath, O., 1929, ``The Scientific Conception of the World: The Vienna Circle'', {\it Wissendchaftliche Weltausffassung}.

\bibitem{Cassini17} Cassini, A., 2016, ``El problema interpretativo de la mec\'anica cu\'antica. Interpretaci\'on minimal e interpretaciones totales'', {\it Revista de Humanidades de Valpara\'iso}, {\bf 8}, 9-42.

\bibitem{Chakravartty17} Chakravartty, A., 2017, {\it Scientific ontology: Integrating naturalized metaphysics and voluntarist epistemology.}, Oxford University Press, New York.

\bibitem{Cordero14} Cordero, N.L., 2014, {\it Cuando la realidad palpitaba}, Biblos, Buenos Aires.

\bibitem{PS} Curd, M. $\&$ Cover, J. A., 1998, {\it Philosophy of Science. The central issues}, Norton and Company (Eds.), Cambridge University Press, Cambridge.


\bibitem{deRonde16b} de Ronde, C., 2016, ``Representational Realism, Closed Theories and the Quantum to Classical Limit'', in {\it Quantum Structural Studies}, pp. 105-135, R.E. Kastner, J. Jeknic-Dugic and G. Jaroszkiewicz (Eds.), World Scientific, Singapore.

\bibitem{deRonde17} de Ronde, C., 2017, ``Causality and the Modeling of the Measurement Process in Quantum Theory'', {\it Disputatio}, {\bf 9}, 657-690.



\bibitem{deRonde20} de Ronde, C., 2020, ``The (Quantum) Measurement Problem in Classical Mechanics'',  in {\it Probing the Meaning of Quantum Mechanics: Entanglement, Correlations and Measurement}, D. Aerts, J. Arenhart, C. de Ronde and G. Sergioli (Eds.), World Scientific, Singapore, in press. (quant-ph:2001.00241).

\bibitem{Deutsch04} Deutsch, D., 2004, {\it The Beginning of Infinity. Explanations that Transform the World}, Viking, Ontario. 

\bibitem{Deutsch11} Deutsch, D., 2011, `Apart from universes'', Talk at the {\it Everett@50 Conference}: https://vimeo.com/5490979

\bibitem{Dirac74} Dirac, P.A.M., 1974, {\it The Principles of Quantum Mechanics}, 4th Edition, Oxford University Press, London.

\bibitem{Einstein65} Einstein, A., 1949, ``Remarks concerning the essays brought together in this co-operative volume'', in {\it Albert Einstein. Philosopher-Scientist}, P.A. Schlipp (Eds.), pp. 665-689, MJF Books, New York. 

\bibitem{EPR} Einstein, A., Podolsky, B. $\&$ Rosen, N., 1935, ``Can Quantum-Mechanical Description be Considered Complete?'', {\it Physical Review}, {\bf 47}, 777-780.

\bibitem{ESPL} Einstein, A., Schr\"odinger, E., Planck M. $\&$ Lorentz H.A., 1967, {\it Letters on Wave Mechanics}. K. Przibram (Ed.), Philosophical Library, New York.

\bibitem{Fine86} Fine, A., 1986, {\it The Shaky Game}, University of Chicago Press, Chicago.

\bibitem{Freire15} Freire Jr., O., 2015, {\it The Quantum Dissidents. Rebuilding the Foundations of Quantum Mechanics (1950-1990)}, Springer, Berlin. 

\bibitem{French11} French, S., 2011, ``Metaphysical underdetermination: Why worry?'', {\it Synthese}, {\bf 180}, 205-221.

\bibitem{FuchsPeres00} Fuchs, C.A. $\&$ Peres A., 2000, ``Quantum theory needs no `interpretation''', {\it Physics Today} {\bf 53}, 70-71.

\bibitem{Hanson} Hanson, R., 1958, {\it Patterns of Discovery: An Inquiry into the Conceptual Foundations of Science}, Cambridge University Press, Cambridge.

\bibitem{Heis58} Heisenberg, W., 1958, {\it Physics and Philosophy}, World perspectives, George Allen and Unwin Ltd., London.

\bibitem{Heis71} Heisenberg, W., 1971, {\it Physics and Beyond}, Harper \& Row, New York.

\bibitem{Heis73} Heisenberg, W., 1973, ``Development of Concepts in the History of Quantum Theory'', in {\it The Physicist's Conception of Nature}, pp. 264-275, J. Mehra (Ed.), Reidel, Dordrecht.

\bibitem{Hofweber16} Hofweber, T., 2016, ``Thomas. Carnap's Big Idea'', in {\it Ontology after Carnap}, pp. 13-30, S. Blatti and S. Lapointe (Eds.), Oxford University Press, Oxford.

\bibitem{Howard04} Howard, D., 2004, ``Who Invented the `Copenhagen Interpretation?' A Study in Mythology'', {\it Philosophy of Science}, {\bf 71}, 669-682.

\bibitem{QTen} Jones, S., 2008, {\it The Quantum Ten. A Story of Passion, Tragedy and Science.}, Thomas Allen Publishers, Toronto.

\bibitem{Andrei08} Khrennikov, A., 2008, ``Analysis of the role of von Neumann's projection postulate in the canonical scheme of quantum teleportation and main quantum algorithms'', preprint, quant-ph:0805.3258.

\bibitem{Koznjak20} Koznjak, B., 2020, ``Aristotle and Quantum Mechanics: Potentiality and Actuality, Spontaneous Events and Final Causes'',  {\it Journal for General Philosophy of Science}, DOI:10.1007/s10838-020-09500-y.

\bibitem{KuhlmannPietsch12} Kuhlmann, M. $\&$ Pietsch, W., 2012, {\it Philosophy of physics}, Springer, Heidelberg.

\bibitem{Landsman17} Landsman, K., 2017, {\it Foundations of Quantum Theory}, Springer, Heidelberg.

\bibitem{Maudlin19} Maudlin, T., 2019, {\it Philosophy of Physics. Quantum Theory}, Princeton University Press, Princeton.

\bibitem{Maudlin19b} Maudlin, T., 2019, ``The Problem With Quantum Theory - Full Interview - Tim Maudlin'', {\it The Institute of Art and Ideas}, https://www.youtube.com/watch?v=hC3ckLqsL5M.

\bibitem{Mermin12} Mermin, D., 2012, ``Quantum Mechanics: Fixing the Shifty Split'', {\it Physics Today}, 65, 8-10.

\bibitem{Peres93} Peres, A., 1993, {\it Quantum Theory: Concepts and Methods}, Kluwer Academic Publishers, Dordrecht.

\bibitem{Popper63} Popper, K.R., 1963,  {\it Conjectures and Refutations: The Growth of Scientific Knowledge}, Routledge Classics, London. 

\bibitem{Popper92} Popper, K.R., 1992, {\it The Logic of Scientific Discovery}, Routledge, New York. 

\bibitem{Schlosshauer11} Schlosshauer, M. (Ed.), 2011, {\it Elegance and Enigma. The Quantum Interviews}, Springer-Verlag, Berlin.

\bibitem{VF80} Van Fraassen, B.C., 1980, {\it The Scientific Image}, Clarendon, Oxford.

\bibitem{VF91} Van Fraassen, B.C., 1991,  {\it Quantum Mechanics: An Empiricist View}, Clarendon, Oxford.

\bibitem{Vernant06} Vernant, J.-P., 2006, {\it Myth and Thought among the Greeks}, Mariner books, New York.

\bibitem{VN} Von Neumann, J., 1996, {\it Mathematical Foundations
of Quantum Mechanics}, Princeton University Press (12th. edition),
Princeton.

\bibitem{Winther16} Winther, R.G., 2016, ``The Structure of Scientific Theories'', {\it The Stanford Encyclopedia of Philosophy (Winter 2016 Edition)}, E. N. Zalta (Ed.), URL: https://plato.stanford.edu/archives/win2016/entries/structure-scientific-theories/.

\bibitem{Raoni20} Wohnrath Arroyo, R., 2020, {\it Discussions on physics, metaphysics and metametaphysics: Interpreting quantum mechanics}, PhD dissertation, Universidad Federal de Santa Catarina. 


\end{thebibliography}
\end{document}